# Potentials with partly constant FSR: semiclassical theory and applications to SNAP microresonators

**M. SUMETSKY**

*Aston Institute of Photonics Technology, Aston University, Aston Triangle, Birmingham, UK, B47ET*
*m.sumetsky@aston.ac.uk*

**Propagation of whispering gallery modes in Surface Nanoscale Axial Photonics (SNAP) microresonators, fabricated at the optical fiber surface, is commonly described by a one-dimensional wave equation, resembling the Schrödinger equation, where the fiber cutoff frequency (CF) varying along the fiber length plays the role of potential and the light frequency plays the role of energy. Of particular importance for applications including frequency comb generation, frequency conversion, and signal processing are SNAP microresonators with constant free spectrum range (FSR). Here we note that, in addition to CF potentials with a globally constant FSR, the potentials having constant FSR confined within a specific spectral region may be sufficient or, in certain cases, preferable for a range of applications. We describe such potentials in semiclassical approximation and analyze their properties considering representative examples.**

In optical and photonic resonant systems, ranging from the macroscopic laser cavities [1] to microresonators [2-7], the request for constant FSR is motivated by the need to suppress dispersion-induced distortions and maintain phase coherence across multiple resonances. For example, in frequency comb generation, a constant FSR translates the input coherent light to evenly spaced frequency lines, enabling precise metrology and spectroscopy [7]. In delay lines and signal processors, uniform spacing prevents pulse broadening [8] and enhances the precision of transitions between harmonically modulated resonator modes [9, 10].

A particularly promising approach to fabricate microresonators with constant FSR is based on Surface Nanoscale Axial Photonics (SNAP) [8, 11], where nanoscale variations in the effective fiber radius are used to localize and manipulate WGMs along an optical fiber. Prospective SNAP devices based on microresonators with constant FSR include miniature delay lines [8, 12], frequency comb generators [13-15], and frequency converters [16].

In the SNAP theory, thanks to dramatically small nanometer-scale variation of the fiber effective radius along its axial coordinate $z$, the expression for a WGM with azimuthal and radial quantum numbers $m$ and $p$ is factorized in cylindrical coordinates $(z, \varrho, \varphi)$ as $E_{mp}(\varphi, r, z, t) =$ $\exp(im\varphi) Q_{mp}(\varrho) \Psi_{mp}(z) \exp(i\omega t)$, where the WGM axial dependence, $\Psi_{mp}(z)$, is described by the equation [11]:

$$\frac{\partial^2 \Psi(z,\omega)}{\partial z^2} + \beta^2(z,\omega)\Psi(z) = 0, \quad (1)$$

$$\beta(z,\omega) = \sqrt{2\mu(\omega - \omega_{cut}(z))}, \quad \mu = \frac{n_0^2 \omega}{c^2}. \quad (2)$$

Here $\beta(z, \omega)$ is the WGM propagation constant, $c$ is the speed of light, $n_0$ is the unperturbed fiber refractive index, $\omega$ is the frequency of light, and $\omega_{cut}(z)$ is the cutoff frequency corresponding to WGM $E_{mp}$. Since, in the consideration below, we consider the WGM spectrum near a single cutoff frequency only, the indices $m$ and $p$ in Eqs. (1) and (2) are omitted for brevity.

Eq. (1) is identical to the stationary Schrödinger equation in atomic units, where $\omega$, $\omega_{cut}(z)$, and $\mu$ play the role of energy, potential, and mass of a particle, respectively. Having in mind this analogy, below we will call $\omega_{cut}(z)$ the cutoff frequency (CF) potential, or simply the potential. In a SNAP microresonator, the CF potential $\omega_{cut}(z)$ has the form of a quantum well, which possesses eigenmodes $\Psi_q(z)$ with eigenfrequencies $\omega_q$ numbered by the axial quantum number $q$ and FSR $\omega_q - \omega_{q-1} = \Delta\omega_q^{(FSR)}$.

First, we address the problem of partly constant FSR in the WKB semiclassical (SC) approximation assuming that the potential well $\omega_{cut}(z)$ is smooth enough. In this approximation, we have:

$$\Psi_q(z) \cong \Psi_q^{(SC)}(z) = \frac{1}{\sqrt{\beta(z,\omega_q)}} \cos\left(\int_{z_1(\omega_q)}^{z} \beta(z,\omega_q)dz - \frac{\pi}{4}\right), \quad (3)$$

where the frequency eigenvalues are determined by the quantization condition

$$\int_{z_1(\omega_q)}^{z_2(\omega_q)} \beta(z,\omega_q)dz = \pi\left(q + \frac{1}{2}\right), \quad q \gg 1, \text{ integer.} \quad (4)$$

Here $x_j(\omega)$ are turning points (zeros of propagation constant $\beta(z, \omega)$). In the semiclassical approximation, the FSR of an optical microresonator, $\Delta\omega_q^{(FSR)}$, is simply expressed through the period of classical oscillations (CO), $T_{CO}(\omega)$, and the CO frequency, $\Delta\omega_{CO}(\omega) = 2\pi/T_{CO}(\omega)$. Recalling the known expression for $T_{CO}(\omega)$ [17], it is straightforward to find by perturbation of Eqs. (4) and (2), that, in the semiclassical approximation, the FSR is equal to CO frequency,

$$\Delta\omega_q^{(FSR)} \cong \Delta\omega_{CO}(\omega_q), \qquad (5)$$

$$\Delta\omega_{CO}(\omega) = \pi\sqrt{\frac{2}{\mu}}\left(\int_{z_1(\omega)}^{z_2(\omega)}\frac{dz}{\sqrt{\omega-\omega_{cut}(z)}}\right)^{-1} \qquad (6)$$

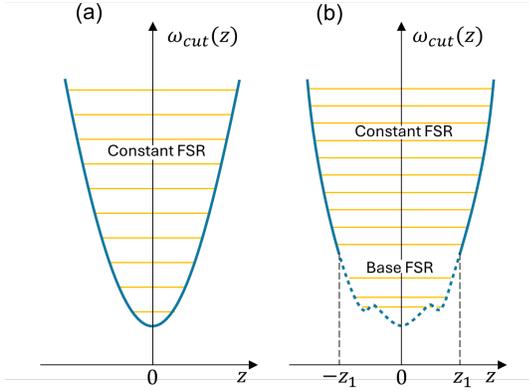

**Fig. 1.** (a) Parabolic potential well. (b) Potential well with known spectrum in its lower base part and constant FSR in the upper part to be designed.

Provided that the required dependence of the FSR $\Delta\omega_q^{(FSR)}$ on the quantum number $q$, and, thus, the dependence of the CO frequency $\Delta\omega_{CO}(\omega)$ on frequency $\omega$, is known, the potential $\omega_{cut}(z)$ is determined by the solution of the Abel integral equation. For simplicity, below we consider *axially symmetric potentials* when $\omega_{cut}(-z) = \omega_{cut}(z)$ (see Fig. 1). Then the solution of Eq. (6) for the branch of $\omega_{cut}(z)$ at $z > 0$ can be expressed through the function inverse to $\omega_{cut}(z)$ as [17]

$$z(\omega_{cut}) = \frac{1}{\sqrt{2\mu}}\left(\int_{\omega_{cut}(0)}^{\omega_{cut}}\frac{d\omega}{\Delta\omega_{CO}(\omega)\sqrt{\omega_{cut}-\omega}}\right), \qquad (7)$$

while for the negative $z$ we use $\omega_{cut}(-z) = \omega_{cut}(z)$. Eqs. (6) and (7), commonly expressed through the period of classical oscillations $T_{CO}(\omega)$ rather than their frequency $\Delta\omega_{CO}(\omega)$, were comprehensively investigated in classical and quantum mechanics, where the problem of *isochronous potentials* (i.e., potentials with frequency independent $T_{CO}(\omega)$, $\Delta\omega_{CO}(\omega)$, and constant FSR $\Delta\omega_q^{(FSR)}$) is of great interest [18-25].

One of the most important examples of isochronous potentials in classical mechanics is the harmonic oscillator when $\omega_{cut}(z) = \omega_0 + \mu\Delta\omega_{CO}^2 z^2/2$ with constant oscillation frequency $\Delta\omega_{CO}$ (Fig. 1(a)). In quantum mechanics, the classical isochronous motion in this potential corresponds to the constant (i.e., independent of quantum number $q$ and frequency) FSR $\Delta\omega_q^{(FSR)} = \Delta\omega_{CO}$ following from Eqs. (1) and (2) as well as from Eqs. (5) and (6). Generally, the latter two equations show that the isochronous potentials possess constant FSR in the semiclassical approximation. This result is well known to be exact for the harmonic potential. However, it is also known that the potentials which are isochronous classically may not possess constant FSR in quantum mechanics, and, conversely, potentials, whose FSR found from the Schrödinger equation is constant, may not be isochronous in classical mechanics [20, 21].

The search for isochronous potentials and potentials with a constant FSR in classical and quantum mechanics has been traditionally focused on *globally isochronous and constant FSR systems*, such as the harmonic oscillator [18-25]. In these idealized cases, the oscillation period (in classical systems) or FSR (in quantum systems) remains uniform across the entire spectral domain. However, to our knowledge, the potentials which are *locally isochronous and/or have constant FSR within a finite bandwidth* remain underexplored.

Remarkably, several practical applications in optics and photonics do not require spectral uniformity across the entire resonator spectrum. Instead, it is often sufficient, and in some cases necessary, to achieve constant FSR over a limited, application-specific bandwidth. For example, limiting the uniform FSR bandwidth of an optical frequency comb generator can concentrate the output power within the desired spectral range, rather than spread it across unnecessary frequencies [26]. In another example, the limitation of the uniform FSR bandwidth is required for the electrooptically generated multiquantum frequency conversion by an optical microresonator system [27, 16].

Our interest in the problem of potentials having a spectrum with *locally constant FSR* was also motivated by the recent experiment reported in Ref. [28], where a SNAP microresonator exhibiting such a spectrum was demonstrated. This experimental observation leads naturally to the theoretical problem illustrated in Fig. 1(b). We assume below that the base of a potential well at $|z| \leq z_1$ (dashed in Fig. 1(b)) as well as the distribution of the CO frequency $\Delta\omega_{CO}(\omega)$ at $|z| \leq z_1$ are known. The problem consists in finding the potential profile $\omega_{cut}(z)$ of the upper part of this well at $|z| > z_1$ (blue solid curves in Fig. 1(b)) to arrive at the required constant FSR $\Delta\omega_{CO}(\omega) \equiv \Delta\omega_{CO}^{(0)}$ within this part, i.e., for $\omega > \omega_{cut}(z_1)$. Under these conditions, the integral in Eq. (7) can be presented as a sum of the known integral, from 0 to $\omega_{cut}(z_1)$, and the easily calculated integral from $\omega_{cut}(z_1)$ to $\omega_{cut}$ where $\Delta\omega_{CO}(\omega) \equiv \Delta\omega_{CO}^{(0)}$. In particular, we have for $z > z_1$:

$$z(\omega_{cut}) = z_0(\omega_{cut}) + \frac{\sqrt{\omega_{cut}-\omega_{cut}(z_1)}}{\sqrt{2\mu}\Delta\omega_{CO}}, \qquad (8)$$

where the known function

$$z_0(\omega_{cut}) = \frac{1}{\sqrt{2\mu}}\left(\int_{\omega_{cut}(0)}^{\omega_{cut}(z_1)}\frac{d\omega}{\Delta\omega_{CO}(\omega)\sqrt{\omega_{cut}-\omega}}\right). \qquad (9)$$

Thus, in the semiclassical approximation, the problem under consideration is reduced to the solution of the transcendental equation, Eq. (8).

First, we consider the situation when the base part of the potential in Fig. 1(b) at $z \leq z_1$ is a harmonic oscillator with

CO frequency $\Delta\omega_{CO}^{(base)}$ and determine the profile of the rest of this potential $z > z_1$ so that at frequencies $\omega > \omega_{cut}(z_1)$ this potential had the CO frequency equal to $\Delta\omega_{CO}^{(0)}$. Under these conditions, the integral in Eq. (9) can be easily calculated and transcendental Eq. (8) is reduced to a quadratic equation that can be solved analytically. Here, rather than identifying the physically meaningful branches of analytical solution, we investigate the potential $\omega_{cut}(z)$ found from Eq. (8) numerically. In Fig. 2(a) we present a harmonic potential having $\Delta\omega_{CO}^{(0)} = \Delta\omega_{CO}^{(base)}$. In this example, as well as in all other examples shown in Fig. 2, we set $\Delta\omega_{CO}^{(base)}/2\pi = 1$ GHz (left plot of Fig. 2(a)) and the base height equal to $\Delta\omega^{(base)}/2\pi = (\omega_{cut}(z_1) - \omega_{cut}(0))/2\pi = 21$ GHz, which corresponds to $z_1 = 0.489$ mm. In Fig. 2(b), we set $\Delta\omega_{CO}^{(0)} = 0.5$ GHz as specified in the left plot of this figure. It is seen that the corresponding potential profile at $z > z_1$ promptly approaches the parabolic asymptote (dashed yellow curve):

$$\Delta\omega_{cut}^{(par)}(z) = \mu\left(\Delta\omega_{CO}^{(0)}\right)^2 \frac{z^2}{2} + \left(1 - \frac{\Delta\omega_{CO}^{(0)}}{\Delta\omega_{CO}^{(base)}}\right)\Delta\omega^{(base)}. \quad (10)$$

Remarkably, while the potentials having $\Delta\omega_{CO}^{(0)} \leq \Delta\omega_{CO}^{(base)}$ always exists, their profile becomes unphysical if $\Delta\omega_{CO}^{(0)} > \Delta\omega_{CO}^{(base)}$. Fig. 2(c), where we set $\Delta\omega_{CO}^{(0)} = 2$ GHz, illustrates this situation leading to the multivalued potential profiles.

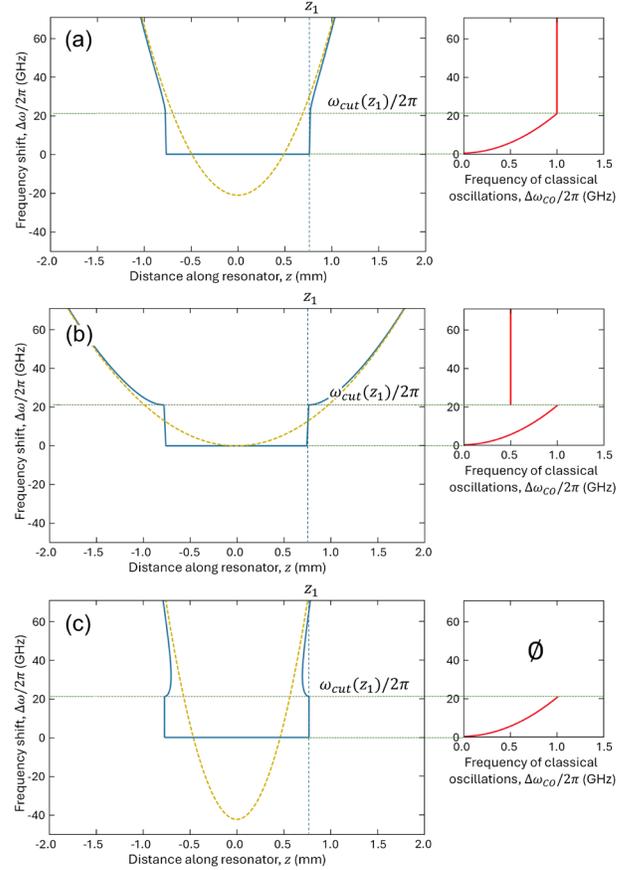

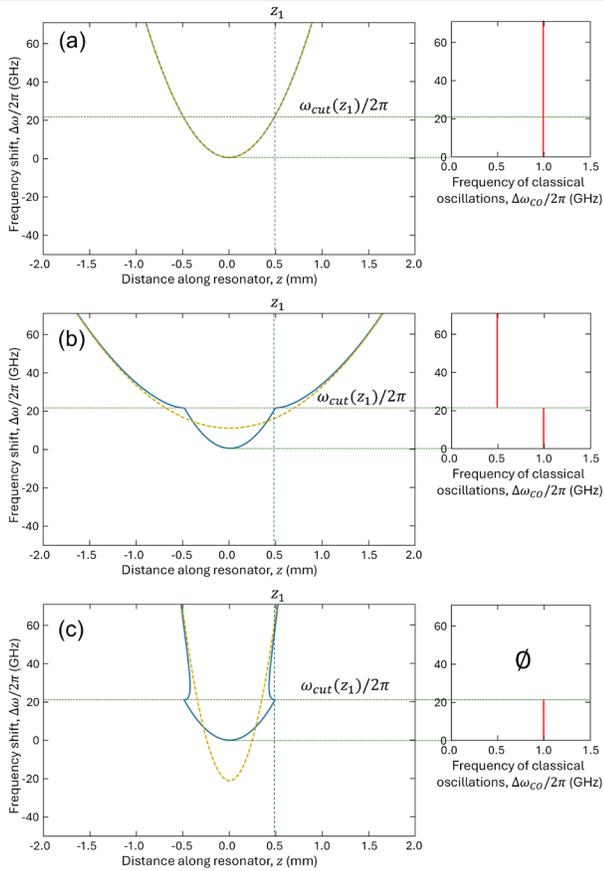

**Fig. 2.** (a) Parabolic potential well with FSR equal to 1 GHz. (b) Potential well with parabolic lower part having the FSR equal to 1 GHz and upper part with $\Delta\omega_{CO}^{(0)} = 0.5$ GHz. (c) The unrealistic profile of potential having the FSR equal to 1 GHz in its lower part and $\Delta\omega_{CO}^{(0)} = 2$ GHz in the upper part. Blue curves are the potential profiles, and yellow dashed curves are their parabolic asymptotes determined by Eq. (10).

**Fig. 3.** (a) Potential well with rectangular lower part whose CO frequency $\Delta\omega_{CO}(\omega)$ matches the constant $\Delta\omega_{CO}^{(0)} = 1$ GHz of its upper part at $\omega = \omega_{cut}(z_1)$. (b) Potential well with rectangular lower part the same as in (a) having the upper part CO frequency equal to $\Delta\omega_{CO}^{(0)} = 0.5$ GHz. (c) The unrealistic profile of potential with rectangular lower part the same as in (a) and (b) having the upper part CO frequency equal to $\Delta\omega_{CO}^{(0)} = 2$ GHz.

Next, in Fig. 3, we consider potentials having a rectangular base, where the CO frequency depends on frequency as $\Delta\omega_{CO}(\omega) = [(\omega - \omega_{cut}(0))/\Delta\omega^{(base)}]^{1/2}\Delta\omega_{CO}^{(base)}$ changing from zero at the base's bottom to $\Delta\omega_{CO}^{(base)} = 1$ GHz at the base's top, which corresponds to $\Delta\omega^{(base)}/2\pi = 21$ GHz and $z_1 = 0.767$ mm. Fig. 3(a) shows the profile of the potential with this base and constant CO frequency $\Delta\omega_{CO}^{(0)}$ in the upper part of the potential which matches the value of CO frequency at the base's top, $\Delta\omega_{CO}^{(0)} = \Delta\omega_{CO}^{(base)} = 1$ GHz. In this plot, as well as in other plots of Fig. 3, the dashed yellow curve is the parabolic asymptote of this potential defined slightly different from that in Eq. (10):

$$\Delta\omega_{cut}^{(par)}(z) = \mu\left(\Delta\omega_{CO}^{(0)}\right)^2 \frac{z^2}{2} + \left(1 - \frac{2\Delta\omega_{CO}^{(0)}}{\Delta\omega_{CO}^{(base)}}\right)\Delta\omega^{(base)}. \quad (11)$$

In Fig. 3(b), we set $\Delta\omega_{CO}^{(0)} = 0.5$ GHz (see the plot at the right hand side of this figure). Similar to the case of potential with a parabolic base shown in Fig. 2, the determined potentials become multivalued and, thus, unphysical at the CO frequencies $\Delta\omega_{CO}^{(0)} > \Delta\omega_{CO}^{(base)}$. This situation is illustrated in Fig. 3(c) for $\Delta\omega_{CO}^{(0)} = 2$ GHz.

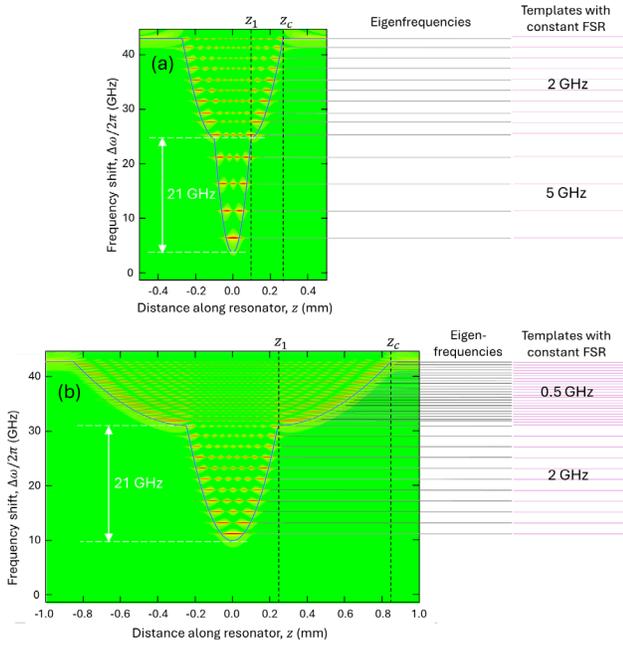

**Fig. 4.** Truncated Green's function of Eq. (1) found for the potential with parabolic base determined from Eqs. (8) and (9) for (a) $\Delta\omega_{CO}^{(0)} = 2$ GHz, $\Delta\omega_{CO}^{(base)} = 5$ GHz and (b) $\Delta\omega_{CO}^{(0)} = 0.5$ GHz, $\Delta\omega_{CO}^{(base)} = 2$ GHz. The determined eigenfrequencies are compared with templates of constant FSRs shown at the right hand side of the figure.

The semiclassical approximation considered so far is valid only for sufficiently smooth potentials and large quantum numbers $q$. For the potentials considered above, this condition is violated near the transition regions for the frequencies $|\omega - \omega_{cut}(z_1)| \lesssim \max(\Delta\omega_{CO}^{(0)}, \Delta\omega_{CO}^{(base)})$ as well as close to the bottom of the rectangular potential in Fig. 3 where $|\omega - \omega_{cut}(0)| \lesssim \Delta\omega_{CO}^{(base)}$. As an example, Fig. 4 presents the results of numerical calculation of the absolute value of the Green's function of Eq. (1) for potential models similar to those considered in Fig. 2, whose bottom part (harmonic oscillator) and upper part are calculated using the classical Eqs. (8) and (9). For better visualization of the regions close to the Green's function poles corresponding to the *frequency eigenvalues*, this function was truncated at a smaller value corresponding to the green area in the plots of Fig. 4. In contrast to potentials considered in Figs. 2 and 3, the potentials shown in Fig. 4 become constant for sufficiently large $|z| > z_c$ and possess continuous spectrum for frequencies $\omega > \omega_{cut}(z_c)$. In Fig. 4(a), the potential profiles are shown for $\Delta\omega_{CO}^{(0)} = 2$ GHz and $\Delta\omega_{CO}^{(base)} = 5$ GHz, while in Fig. 4(b) we set $\Delta\omega_{CO}^{(0)} = 0.5$ GHz and $\Delta\omega_{CO}^{(base)} = 2$ GHz. It is seen that the accurately calculated FSRs deviate from their semiclassical values near the transition region $|\omega - \omega_{cut}(z_1)| \lesssim \max(\Delta\omega_{CO}^{(0)}, \Delta\omega_{CO}^{(base)})$ and close to continuous spectrum and match their semiclassical values much better away from these regions. Obviously, increasing the potential height and axial length will lead to increased numbers of supported eigenmodes and a better precision of semiclassical approximation.

In conclusion, inspired by the recent experiment observation of SNAP microresonators with partly constant FSR [28], we developed the semiclassical theory and presented examples of such microresonators described by CF potentials illustrated in Fig. 1(b). The considered examples include potentials with parabolic and rectangular base. We showed that it is possible to design CF potentials whose partly constant FSR is smaller than the FSR of the CF potential base, while the potentials having a larger constant FSR do not exist. The determined semiclassical solutions are in a reasonable agreement with accurate numerical solutions of the wave equation. We suggest that the results obtained may be useful for the design of frequency comb generators, frequency converters, and signal processors based on SNAP microresonators.

**Funding.** Engineering and Physical Sciences Research Council (EPSRC) grants EP/W002868/1 and EP/X03772X/1, Leverhulme Trust grant RPG-2022-014.

**Disclosures.** The authors declare no conflicts of interest.

**Data availability.** Data underlying the results presented in this paper may be obtained from the corresponding author upon reasonable request.

### References

1. A. E. Siegman, *Lasers* (University science books, 1986).
2. K.J. Vahala, "Optical microcavities," Nature **424**, 839 (2003).
3. A.B. Matsko (Ed), Practical Applications of Microresonators in Optics and Photonics (CRC Press, 2009).
4. G.C. Righini, et al., "Whispering gallery mode microresonators: fundamentals and applications," Riv. Nuovo Cimento **34**, 435 (2011).
5. D.V. Strekalov, et al., "Nonlinear and quantum optics with whispering gallery resonators," J. Opt. **18**, 123002 (2016).
6. D. Yu, et al., "Whispering-gallery-mode sensors for biological and physical sensing," Nat. Rev. Methods Primers, **1**, 83 (2021).
7. Y. Sun, et al., "Applications of optical microcombs," Adv. Opt. Photon. **15**, 86 (2023).
8. M. Sumetsky, "Delay of Light in an Optical Bottle Resonator with Nanoscale Radius Variation: Dispersionless, Broadband, and Low Loss," Phys. Rev. Lett. **111**, 163901 (2013).
9. A. Parriaux, K. Hammani, and G. Millot, "Electro-optic frequency combs," Adv. Opt. Photon. **12**, 223 (2020).
10. M. Zhang, et al., "Integrated lithium niobate electro-optic modulators: when performance meets scalability," Optica **8**, 652 (2021).
11. M. Sumetsky, "Theory of SNAP devices: basic equations and comparison with the experiment," Opt. Express **20**, 22537–22554 (2012).
12. N. Toropov, et al., "Microresonator devices lithographically introduced at the optical fiber surface," Opt. Lett. **46**, 1784 (2021).
13. S.V. Suchkov, M. Sumetsky, and A.A. Sukhorukov, "Frequency comb generation in SNAP bottle resonators," Opt. Lett. **42**, 2149 (2017).
14. Y.V. Kartashov, et al., "Two-dimensional nonlinear modes and frequency combs in bottle microresonators," Opt. Lett. **43**, 2680 (2018).

15. V. Dvoyrin, and M. Sumetsky, "Bottle microresonator broadband and low-repetition-rate frequency comb generator," Opt. Lett. **41**, 5547 (2019).
16. M. Sumetsky, "Complete inelastic transparency of time modulated resonant photonic circuits," in Proc. SPIE PC12912, Quantum Sensing, Imaging, and Precision Metrology II, PC1291215 (13 March 2024); https://doi.org/10.1117/12.3011746.
17. L. D. Landau, and E. M. Lifshitz, *Mechanics* (Elsevier, 1982).
18. F. Calogero, *Isochronous systems* (OUP Oxford, 2008).
19. E. T. Osypowski and M. G. Olsson, "Isochronous motion in classical mechanics," Am. J. Phys. **55**, 720 (1987).
20. M. Asorey, et al., "Isoperiodic classical systems and their quantum counterparts," Ann. Phys. **322**, 1444 (2007).
21. J. F. Carinena, et. al., "Isochronous classical systems and quantum systems with equally spaced spectra," J. Phys.: Conf. Series **87**, 012007, (2007).
22. G. Gorni and G. Zampieri, "Global isochronous potentials," Qualitative theory of dynamical systems, **12**, 407 (2013).
23. D. J. Cross, "Every isochronous potential is shear-equivalent to a harmonic potential," Amer. J. Phys. **86**, 198 (2018).
24. F. Teichert, E. Kuhn, and A. Thränhardt, "Do equidistant energy levels necessitate a harmonic potential?" Opt. Quant. Electron. **53**, 403 (2021).
25. D. V. Treschev, "On isochronicity," Proceedings of the Steklov Institute of Mathematics, **322**, 198 (2023).
26. B. Buscaino, et al., "Design of efficient resonator-enhanced electro-optic frequency comb generators," J. Lightwave Technol. **38**, 1400 (2020).
27. Y. Hu, et al., "On-chip electro-optic frequency shifters and beam splitters," *Nature* **599**, 587 (2021).
28. I. Sharma and M. Sumetsky, "Widely FSR tunable high Q-factor microresonators formed at the intersection of straight optical fibers," arXiv preprint arXiv:2504.10364 (2025).